\documentclass[prl,twocolumn,showpacs]{revtex4}
\usepackage{epsfig}
\begin{document}
\title{
Stochastic Ergodicity Breaking: a Random Walk Approach.
}
\author{G. Bel, E. Barkai}
\affiliation{Department of Physics,
Bar Ilan University,
Ramat-Gan 52900 Israel}
\begin{abstract}
{ The continuous  time random walk (CTRW) model exhibits a non-ergodic phase
when the average waiting time diverges. Using an analytical approach
for the non-biased and the uniformly biased CTRWs, and numerical simulations
for the CTRW in a potential field, 
we obtain the non-ergodic properties of the random
walk which show strong deviations from Boltzmann--Gibbs theory.
We derive 
 the distribution
function of occupation times  in a bounded region of space
which, in 
the ergodic phase recovers the Boltzmann--Gibbs theory,
while in the non-ergodic phase yields
a generalized non-ergodic statistical law. 
}
\end{abstract}
\pacs{02.50.-r,05.40.-a,02.50.Ey}

\maketitle

 The ergodic hypothesis is a cornerstone of
statistical mechanics. It states that  ensemble
averages  and time averages are equal
in the limit of infinite measurement
time. 
Starting with the work of Bouchaud, there is growing interest
in stochastic ergodicity breaking (SEB)  
which finds applications in a wide range of 
physical systems: phenomenological models of glasses \cite{Werg},
laser cooling \cite{Bardou,Bardou1}, blinking
quantum  dots \cite{Brokmann,Gennady2}, 
and models of atomic transport in optical lattices \cite{Lutz}. 
SEB is found for systems,
whose dynamics is characterized by power law sojourn times,
with infinite average waiting times. 
In such systems the microscopical time scale diverges,
for example the average trapping time of an atom in the
theory of laser cooling \cite{Bardou,Bardou1}. The relation
between SEB and diverging sojourn times
can be briefly explained, by noting that one condition
to obtain ergodicity is that the measurement time is long,
compared with the characteristic time scale of the problem.
However this condition is never fulfilled
if  the microscopical time scale,  i.e. the average trapping
time, is infinite.
It is important
to note that the concept of trapping time probability density function
(PDF) $\psi(t)$,
with diverging first moment, is wide spread and found in
many fields of physics \cite{Bardou,STR,Hughes,Metzler}. 
It was introduced into physics by Scher and Montroll
in the context of continuous
time random walk (CTRW) \cite{Montroll}. 
This well known model \cite{Hughes,Metzler}
exhibits anomalous
sub-diffusion $\langle r^2 \rangle\sim t^{\alpha}$ with $\alpha<1$, and aging behaviors \cite{Barkai}
which are related to
SEB.

Clearly if the CTRW is non-ergodic, Boltzmann--Gibbs statistics
is not valid, in a way defined precisely later. 
The goal of this letter is to obtain a generalization of 
Boltzmann--Gibbs statistical mechanics, for CTRW models.
Besides its theoretical importance this 
goal is timely due to recent observations
on the single particle level of CTRW type of dynamics 
\cite{Xie,Weitz}, 
for
example anomalous diffusion of 
a single magnetic bead in a polymer network
with a well defined temperature $T$ \cite{Weitz}. In single
particle experiments, the many particle averaging, i.e. the
problem of ensemble averaging, is removed \cite{BarRev}.
Hence a fundamental question is whether
time averages of single particle trajectories yield
information identical to ensemble averages. 
The large number of applications of the CTRW
model, and related models like the trap model and the comb model,
make us believe that constructing a general non-ergodic theory
for such systems is worthy.

 Before introducing the model, recall that 
the  basic tool in statistical mechanics is Boltzmann's   
probability $P_B(x)$ of finding a system  
in a  state with energy $E(x)$, using the canonical ensemble
\begin{equation}
P_{B}(x) = {\exp\left( - {E(x) \over T} \right) \over Z},
\label{EqBol}
\end{equation}
where $T$ is the temperature and $Z= \sum_x \exp[ -E(x)/T]$
is the normalizing partition function. Here
for simplicity we assume a discrete energy spectrum.
To obtain the {\em  ensemble} average of a physical observable,
for example the
energy of the system, we use
$ \langle E \rangle = \sum_x E(x) P_{B}(x)$,
and similarly for other physical observables  like entropy, free energy  etc. 
 When
measurement of a single system is made, a time average
is recorded. 
Consider a system randomly changing between its energy
states $\{ E(x) \}$. 
Let $t_x$ be the total time spent by the
system in energy state $E(x)$, within the total observation
interval $(0,t)$. 
We define the occupation fraction
$$
\overline{p}_x = { t_x \over t},
\mbox{\ and the time average energy is \ } 
\overline{E} = \sum_x E(x) \overline{p}_x . 
$$
According to statistical mechanics, once the ergodic hypothesis
is satisfied, and within the canonical 
formalism $\overline{p}_x = P_{B}(x)$
and then $\overline{E} = \langle E \rangle$, and similarly
for other physical observables. Thus for ergodic systems
the fraction of occupation time is non-random, in the thermodynamic
limit of long measurement time \cite{Majumdar}. 
For non-ergodic systems the occupation fraction $\overline{p}_x$ is
a random variable, even in the long time limit.
Thus, an important goal of the theory 
of SEB is to calculate the distribution function 
of the random variable $\overline{p}_x$. 
We will show below, in the context of CTRW models,
that  a rather general distribution function,  
describes statistical properties of
the fraction of occupation times in the non-ergodic phase.
And
that this distribution function is related
to the partition function of the problem.

 We consider a
one dimensional CTRW on a lattice. The lattice
points are labeled with index $x$ and $x=-L,-L+1,...,0,...L$,
hence the system size is $2 L+ 1$. On each lattice point
we define a probability $0<Q_R(x)<1$ for jumping right,
and a probability for jumping left $Q_L(x)=1-Q_R(x)$.
Let $\psi(t)$ be the PDF of waiting times at the sites.
The particle starts at site $x=0$, it will wait there
for a period $t_1$ determined from $\psi(t)$, it will
then jump with probability $Q_L(0)$ to the left,
and with probability $Q_R(0)$ to the right. After the
jump, say to lattice point $1$, the particle will 
pause for a period $t_2$, whose statistical properties
is  determined by $\psi(t)$. It will then jump
either back to point $x=0$ or to $x=2$, according to the probability law
$Q_{R}(1)$. 
Then the process is renewed. 
We assume reflecting boundary
conditions, namely $Q_L(L)=Q_R(-L)=1$.
We consider the generic case \cite{Hughes,Metzler}, where 
\begin{equation}
\psi(t) \sim  {A t^{ - ( 1 + \alpha)} \over | \Gamma\left( - \alpha\right) |}
\label{eqLT}
\end{equation}
when $t \to \infty$ and $0<\alpha<1$, $A>0$.
Specific values of $\alpha$   
for a wide range of physical systems and models are given
in \cite{STR,Hughes,Metzler,Montroll,Barkai,Weitz}.
 In this case the average waiting
time is infinite.  

\begin{figure}
\begin{center}
\epsfxsize=70mm
\epsfbox{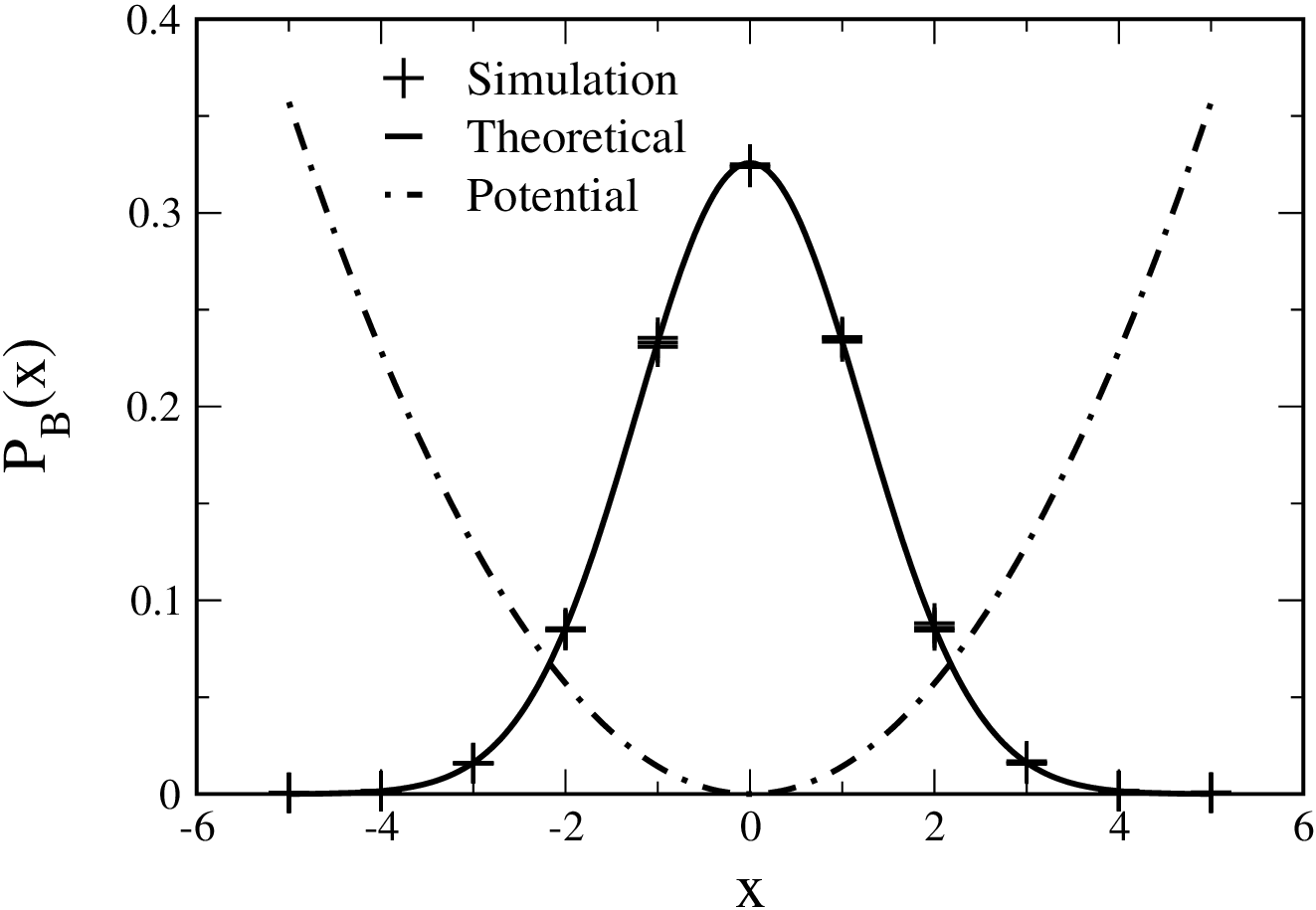}
\end{center}
\caption
{ Boltzmann's equilibrium for an ensemble of CTRW  particles,
in an harmonic potential field, and fixed temperature $T=3$.
In simulations (cross) we use  $\alpha=0.3,0.5,0.8$, 
the results being indistinguishable.
The figure illustrates
that for an ensemble of particles standard equilibrium is obtained.
Ergodicity breaking is found only when long time averages
of single particle trajectories are analyzed.
The scaled potential (dot dash curve) is  the harmonic
potential field, and the theoretical curve is Boltzmann's
equilibrium distribution. 
}
\label{figB1}
\end{figure}

We obtain now an important
limiting distribution function,  which will yield the statistical
properties of the non-ergodic CTRW.
We consider a specific lattice point $x$.
We introduce a state function $\theta_x(t)$ which is equal
$1$ when the particle is on $x$, other wise it is zero.
Thus $\theta_x(t)$ jumps between the value $1$ (state $+$) and zero
(state $-$) and vice versa.
The PDF  of times when the particle occupies  state $+$ [$-$] is denoted with
$\psi_{+}(t)$ $[\psi_{-}(t)]$ respectively. In our model $\psi_{+}(t)=\psi(t)$.
To obtain  $\psi_{-}(t)$ note that  after the particle leaves lattice point $x$ it is either on 
$x+1$ or on $x-1$. 
Let $t_R$ $(t_L)$ be the random time it takes the particle starting
on $x+1$ $(x-1)$ to return to $x$,
and $f_R(t_R)$ [$f_L(t_L)$] the corresponding PDF of the first passage time 
respectively. Then the PDF of times
in state $-$ is
\begin{equation}
\psi_{-}\left(t\right) = Q_R(x) f_R \left(t\right) 
 + Q_L(x) f_L \left(t\right). 
\label{eqsum}
\end{equation}
We will later find explicit expressions
for $\psi_{-}(t)$, however for the time being let us
assume that in the limit of long  $t$ it behaves
like
\begin{equation}
\psi_{-}(t) \sim {A_{x} t^{- (1 + \alpha)} \over |\Gamma\left(-\alpha\right)|},
\end{equation}
where $A_x>0$ will depend on model parameters.
 Let $t_{x}$ be the total time spent on point $x$,
within the time period $(0,t)$. Then the occupation
fraction is
$
\overline{p}_{x} = t_{x}/ t = \int_0 ^t \theta_x(t') {\rm d} t'/ t
$.
A calculation, whose details will be given elsewhere, shows that
the PDF  of the occupation fraction in the limit of infinite measurement time
is
\begin{equation}
 f(\overline{p}_x )= 
 {\rm \delta}_{\alpha} ({\cal R}_x,\overline{p}_{x}),
\end{equation}
where
${\cal R}_x=A / A_x$ 
and 
$ {\rm \delta}_{\alpha}({\cal R}_x,\overline{p}_x) \equiv$
\begin{equation}
{ \sin \pi \alpha \over \pi } 
{{\cal R}_x \overline{p}_{x} ^{\alpha -1} \left( 1 - \overline{p}_{x} \right)^{\alpha-1} \over
{\cal R}_x\ ^2 \left( 1 - \overline{p}_{x} \right)^{2 \alpha} + \overline{p}_{x} ^{2 \alpha} + 2
{\cal R}_x \left( 1 -  \overline{p}_{x} \right)^{\alpha} \overline{p}_{x} ^\alpha \cos \pi \alpha }.
\label{eqGA11}
\end{equation}
This equation indicates SEB since the occupation fraction
remains a random variable, even in the limit of long measurement
times. 
The  PDF 
Eq. (\ref{eqGA11}) is normalized
according to 
$\int_0 ^1 {\rm \delta}_{\alpha}({\cal R}_x,\overline{p}_{x}) {\rm d}  \overline{p}_{x}=1$.
When  ${\cal R}_x=1$, Eq. (\ref{eqGA11}) was obtained by  Lamperti 
\cite{Lamp} in the context of the mathematical theory  of occupation
times \cite{Godreche}, and see \cite{Gennady2}
for a physical application in the context of
blinking quantum dots. 
In particular when ${\cal R}_x=1$ and $\alpha =1/2$ we find
the arcsine distribution. 
More generally the amplitude ratio ${\cal R}_x$ determines the degree of
symmetry in the problem as we will demonstrate later. 

We use a  general  physical argument to obtain ${\cal R}_x$. Assume
that the random walker is in contact with a thermal heat bath, with
temperature $T$, and interacting with
an external potential  field $E(x)$. For an {\em ensemble} of particles
Boltzmann Gibbs statistics must hold. In particular the probability
that a single member of an ensemble of particles, 
will occupy the lattice point
$x$, is given by
$P_B(x)$ Eq. (\ref{EqBol}).
%
%
Eq. 
(\ref{eqGA11}) shows that 
$\overline{p}_x$ is a random variable, 
however when  we ensemble average
the occupation fraction we must obtain Boltzmann-equilibrium
statistics
\begin{equation}
\langle \overline{p}_x \rangle = \int_0 ^1 \overline{p}_x  f\left( \overline{p}_x \right) {\rm d} \overline{p}_x=
P_B(x).
\end{equation}
On the other hand Eq. (\ref{eqGA11}) yields 
\begin{equation}
\langle \overline{p}_x \rangle = { {\cal R}_x \over 1 + {\cal R}_x }.
\end{equation}
Hence we find
\begin{equation}
{\cal R}_x = {P_B (x) \over 1 -P_B(x) }={\exp\left( - {E(x)\over T}\right) \over Z'},
\label{eqRx}
\end{equation}
where $Z'=\sum_{y} ^{'}\exp( - E(y)/T)$ and  
the sum is over all energy states
excluding state $x$.
Eqs. 
(\ref{eqGA11}, 
\ref{eqRx}) 
describe the relation between the non-ergodic dynamics
and the partition function
of the problem. Thus we established an explicit relation between
SEB and the basic tool of equilibrium
statistical mechanics.
The remaining  goal of this Letter is to prove
our physical picture, based on the CTRW model.

\begin{figure}
\begin{center}
\epsfxsize=70mm
\epsfbox{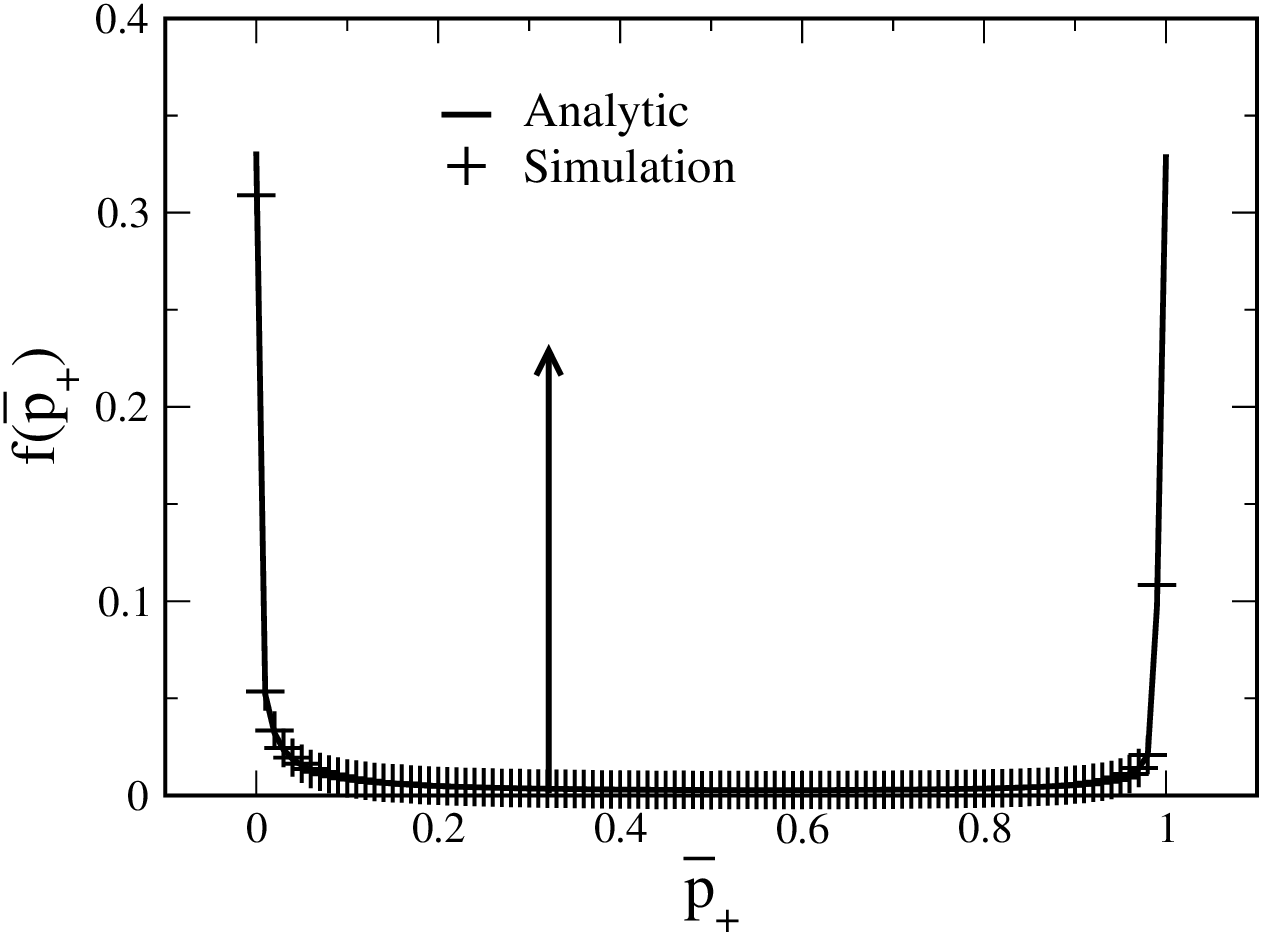}
\end{center}
\caption
{
The PDF of occupation fraction $\overline{p}_x = t_{x}/t$ where
$t_{x}$ is the occupation time
on lattice point $x=0$. The random walk is in an harmonic potential
field, the point $x=0$ being the minimum of energy.
For an ergodic process satisfying detailed
balance, the PDF $f(\overline{p}_x)$ would be narrowly centered around the value
predicted by Boltzmann which is given by the arrow. In a given numerical
 experiment, it is unlikely to obtain
the value of $\overline{p}_x$  predicted using  Boltzmann--Gibbs ergodic theory.
The solid curve is the analytical formula Eq.
(\ref{eqGMR06}) with $\alpha=0.3$ and $T=3$
}
\label{fig304}
\end{figure}

\begin{figure}
\begin{center}
\epsfxsize=70mm
\epsfbox{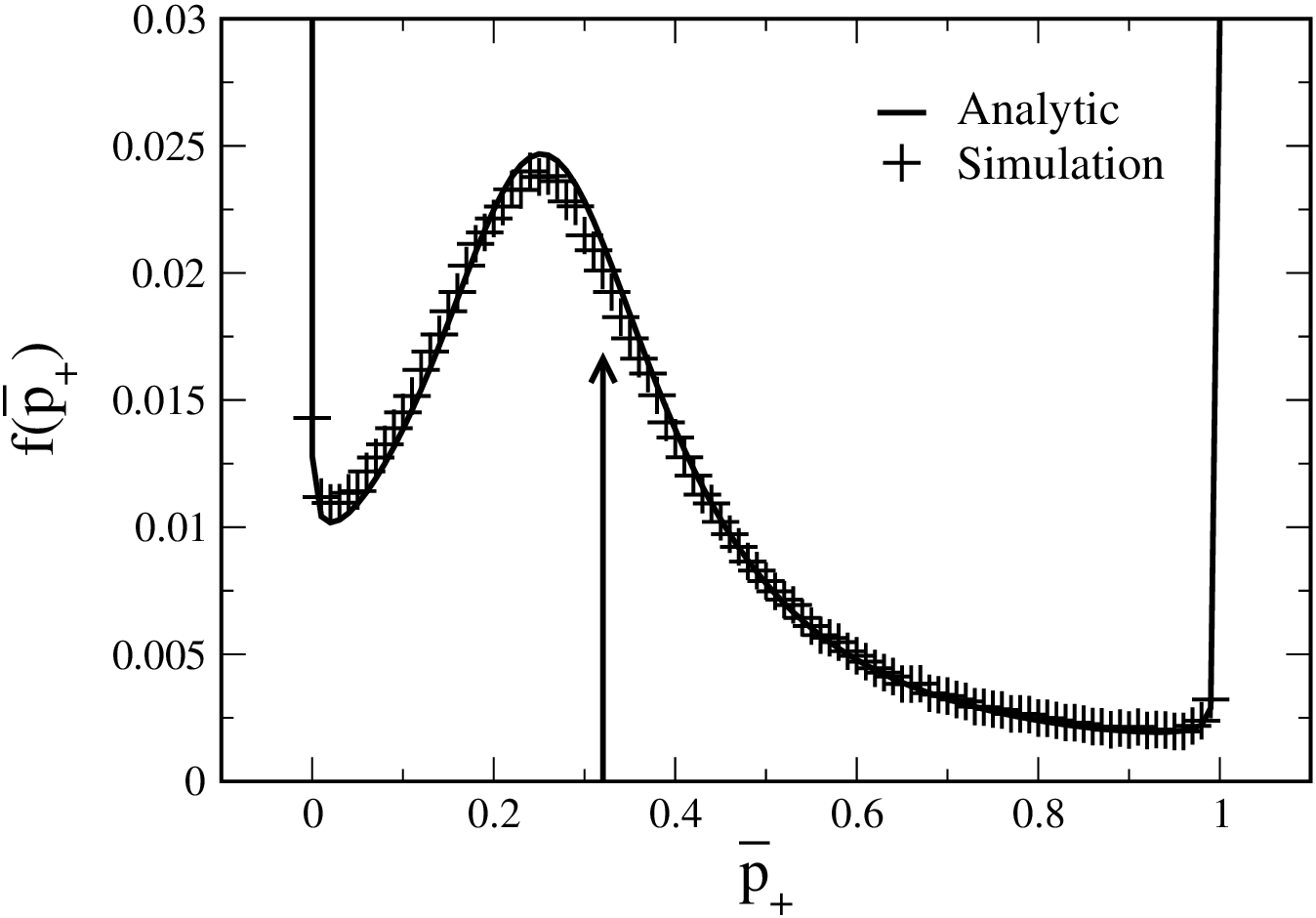}
\end{center}
\caption
{
Same as Fig. (\ref{fig304}) however now $\alpha=0.8$. 
Instead of the $U$ shape found in Fig. (\ref{fig304}) we
find a distorted $W$ shape of the  PDF.
A peak close to Boltzmann's value for $\overline{p}_x$, i.e the arrow
on $P_B(x)$ is an indication
that as $\alpha$ is increased
the ergodic phase is approached. 
}
\label{fig806}
\end{figure}

 We now consider the unbiased one dimensional CTRW on a lattice $x=-L,\cdots,L$,
with $Q_L(x)=Q_R(x)=1/2$. Such a process describes sub-diffusive motion. 
We obtained the long time behaviors of  PDFs $f_R(t)$ and $f_L(t)$  (details
are left for longer publication),
and then obtained  the PDF of the fraction of occupation time
$\overline{p}_{x}$ on a lattice point $x$, excluding the boundary points, 
\begin{equation}
\lim_{t \to \infty} f \left( \overline{p}_{x} \right) = {\rm \delta}_{\alpha}\left(\left(2 L -1\right)^{-1},\overline{p}_{x}\right).
\label{eqGMR00}
\end{equation}
Eq. (\ref{eqGMR00}) does not depend on the position $x$ of the observation point,
reflecting the symmetry of the problem.


 The biased CTRW yields anomalous diffusion with a drift.
For this case the probability of jumping left
is $Q_L(x)=q$, and right is $1-q$ where $q\ne 1/2$. 
 Unlike the unbiased case now clearly different locations along the lattice,
have different distributions of the fraction of occupation
time.
A detailed calculation of the first passage
times,  shows that in the long time limit,
\begin{equation}
f \left( \overline{p}_{x} \right) ={\rm \delta}_{\alpha}\left[ {\cal R}_x, \overline{p}_{x} \right] 
\label{eqGMRo2}
\end{equation}
with
$ {\cal  R}_x= $
\begin{equation}
\left\{ {2 \over 2 q - 1} \left[ q^2 \left( {q \over 1 - q } \right)^{L + x -1} - 
\left( 1 - q \right)^2 \left( { 1 - q \over q} \right)^{L-x-1}\right] - 1\right\}^{-1}.
\label{eqGMR03}
\end{equation}
Note that Eq. (\ref{eqGMRo2}) is not sensitive
to the 
short time behavior of the waiting time
distribution. When $q=1/2$ we recover
Eq. (\ref{eqGMR00}).

%

The biased CTRW is used to model anomalous diffusion
under the influence of a constant
external driving force ${\cal F}$, e.g. \cite{Montroll}. If the physical process
is close to thermal equilibrium,  the condition of
detailed balance is imposed on the dynamics. This  standard condition implies
that for {\em an ensemble of particles}  Boltzmann's equilibrium  is obtained.  
The potential energy at each point $x$, excluding the reflecting
boundaries, due to the
interaction with the external driving force is  
$E(x)=- {\cal F} a x$ and $a$ is the lattice spacing. 
The condition of detailed balance then
reads
\begin{equation}
q={1\over 1 + \exp\left( {{\cal  F} a \over T}\right)} .
\label{eqDB}
\end{equation}
Using Eqs. (\ref{eqGMR03},\ref{eqDB}) we can rewrite the solution
in an elegant form
\begin{equation}
f \left( \overline{p}_{x} \right) ={\rm \delta}_{\alpha} \left({P_B (x) \over 1 - P_B(x) }, \overline{p}_{x} \right),  
\label{eqGMR06}
\end{equation}
where
$P_B \left( x \right)$ is 
the canonical Boltzmann factor 
Eq. (\ref{EqBol}).
When the external force is zero we have $P_B(x)=1/Z$ and $Z=2 L$.
Eq. (\ref{eqGMR06}) proves that our physical arguments
leading to Eq. (\ref{eqRx}) are valid,
at-least for the uniformly biased and unbiased random walks.
More generally, for CTRWs far from thermal equilibrium,
Eq. (\ref{eqGMR06}) is still valid however one must
replace $P_B(x)$ with the corresponding equilibrium   
probability.  

Eq. (\ref{eqGMR06}) 
shows that  the fluctuations $\overline{p}_x$ 
are vast, in particular in any single measurement we are not likely
to measure the averaged value $P_B(x)$ (see details below). 
In the limit $\alpha \to 1$ our theory reduces to the standard
canonical theory, since then $f\left( \overline{p}_x \right) = \delta[\overline{p}_{x} 
-P_B(x)] $.
Using Eq. 
(\ref{eqGMR06}) we can explain the meaning of the asymmetry parameter 
${\cal R}_x$ in Eq. (\ref{eqRx}).
The numerator of ${\cal R}_x$ is
$P_B(x)$, namely the probability (in ensemble sense) of finding
the system in state $+$ (i.e. $\theta_x(t)=1$) while the denominator
$1 - P_B(x)$ is the probability of finding the particle in state
$-$ (i.e. $\theta_x(t)=0$). This interpretation suggests to us that Eq.  
(\ref{eqGMR06}) has a more general validity, beyond the
biased and non-biased CTRW.

 More generally we define an energy profile for the system
$\{ E_{-L}, E_{-L+1}, ...,E_x,\cdots \}$. The well known detailed balance, 
relates between density of particles at equilibrium $N_{eq}(x)$
at points $x$ and say $x-1$ according to
\begin{equation}
 N_{{\rm eq}}(x) Q_L(x) = N_{{\rm eq}}(x-1)\left[ 1 - Q_L(x-1)  \right]
\label{eqDB1}
\end{equation}
where  
$N_{{\rm eq}}(x)/N_{{\rm eq}}(x-1) = 
\exp\left(  { E_{x-1} - E_x \over T} \right)$ 
and hence we get the constrain on the transition
probabilities
$ Q_L(x) /[ 1 - Q_L(x-1)]  = \exp\left( - { E_{x-1} - E_x \over T} \right)$. 
%
%
%
%
For a general energy field we postulate that,
if the energy profile yields a Boltzmann--Gibbs ergodic behavior for 
a waiting time distribution with finite moments (e.g., exponential
waiting times), then for the same energy profile and
a long tailed waiting time PDF $\psi(t)$ given in Eq.  
(\ref{eqLT}),
our central Eq. (\ref{eqGMR06}) is still valid. Now $P_B(x)$
depends of course on the specific energy profile under
investigation. 

 We  check numerically the generality of
Eq. (\ref{eqGMR06}), 
using the example of a random walk in an harmonic potential.
The problem of anomalous diffusion in harmonic
field was considered in the context of 
fractional Fokker--Planck equations \cite{FracPRL} and
in single protein experiments  \cite{Xie}.
As a byproduct, our 
work shows that fractional Fokker-Planck equations \cite{Metzler}
can be used to describe density of  many particles and
not time average quantities, in this sense the fractional kinetic framework
is very different than the standard Fokker-Planck
equations. 

 The potential field we choose is $E(x) = K x^2$, with $K= 1 $,
and $T=3$.
Eq. (\ref{eqDB1}) and the symmetry condition
$Q_L(0)=1/2$ define the set of transition probabilities
$\{Q_L(x)\}$
for the problem.
In Fig. \ref{figB1}
we check that our simulations yield Boltzmann equilibrium 
in the Harmonic field for an ensemble of particles.
We then consider one
trajectory at a time. We obtain from the simulations,
the total time $t_x$ spent by a particle on the lattice point $x=0$,
namely at the minimum of the potential,
and then construct histograms of 
the  occupation fraction $\overline{p}_{x}= t_{x}/t$.

 We consider the case  $\alpha=0.3$ 
in Fig. (\ref{fig304})  and 
show an excellent
agreement between our  non-ergodic theory Eq.
(\ref{eqGMR06})
 and numerical
simulations. The figure exhibits a $U$ shaped PDF. To understand this
behavior, note that 
for $\alpha<<1$ we expect that the particle 
will get stuck on one lattice point during a very
long period, which is of the order of the measurement
time $t$.  This trapping point,
can be either the point of observation (e.g. $x=0$ in our simulations)
or some other lattice point. In these cases we expect
to find $\overline{p}_{x} \simeq 1$ or   
$\overline{p}_{x} \simeq 0$, respectively.
Hence the PDF of $\overline{p}_{x}$ has a $U$ shape.
This is a strong
non ergodic behavior, in the sense 
that we have a very small probability
for  finding  occupation fraction close to the value predicted
based on Boltzmann's ergodic theory (the arrow). 

 When we increase $\alpha$ we anticipate a ``more ergodic'' behavior,
in particular in the limit $\alpha \to 1$. An ergodic
behavior means that the PDF of the occupation
fraction $\overline{p}_{x}$ is centered on the Boltzmann's
probability (i.e. the arrows in the Figs). In Fig. 
\ref{fig806}
we set $\alpha=0.8$ and observe
a peak in the PDF of $\overline{p}_{x}$ centered in 
the vicinity of the ensemble average value. Note however
that the PDF $f(\overline{p}_{x})$ still attains its maximum
on $\overline{p}_{x}=0$ and 
$\overline{p}_{x}=1$.


More generally,
Eqs. (\ref{eqGA11}, 
\ref{eqRx}) 
yield the non-ergodic statistical mechanical theory
of the CTRW model, in the sense that our theory gives
the distribution of $\overline{p}_x$, while 
the ergodic Boltzmann--Gibbs theory states  $\overline{p}_x=P_B(x)$. 
Our arguments leading to  
Eqs. (\ref{eqGA11}, 
\ref{eqRx}) are general, hence our theory 
might not be limited to CTRW models.  
The mathematical foundation of the theory is the limit theorem
(\ref{eqGA11})
related to the arcsine law. The physical input is  the anomalous
diffusion exponent $\alpha$. A connection between the non-ergodic
dynamics and the partition function was found, which enables us to
find non-trivial SEB  properties of the underlying
random walk, in particular the random walk in a potential field.  


{\bf Acknowledgments} This work was supported by the
National Science Foundation award No. CHE-0344930,
and the Center for Complexity Science Jerusalem.

\end{document}